\documentclass[journal,twocolumn]{IEEEtran}

\usepackage{mathtools}
\usepackage{algorithm}
\usepackage{algpseudocode}
\usepackage{epsfig,makeidx,color}
\usepackage{amsmath,amssymb,bbm}
\usepackage{cite,graphicx,lipsum}
\usepackage{enumerate}
\usepackage{listings}
\usepackage[switch,pagewise]{lineno}
\usepackage{hyperref}
\usepackage{algorithmicx}
\hypersetup{
        colorlinks = true,
        citecolor=blue,
}
\usepackage{listings}
\usepackage{xcolor}

\definecolor{codegreen}{rgb}{0,0.6,0}
\definecolor{codegray}{rgb}{0.5,0.5,0.5}
\definecolor{codepurple}{rgb}{0.58,0,0.82}
\definecolor{backcolour}{rgb}{0.95,0.95,0.92}

\lstdefinestyle{mystyle}{
    backgroundcolor=\color{backcolour},   
    commentstyle=\color{codegreen},
    keywordstyle=\color{magenta},
    numberstyle=\tiny\color{codegray},
    stringstyle=\color{codepurple},
    basicstyle=\ttfamily\footnotesize,
    breakatwhitespace=false,         
    breaklines=true,                 
    captionpos=b,                    
    keepspaces=true,                 
    numbers=left,                    
    numbersep=5pt,                  
    showspaces=false,                
    showstringspaces=false,
    showtabs=false,                  
    tabsize=2
}

\def\be{ \begin{equation} }
\def\ee{ \end{equation} }
\def\bea{ \begin{eqnarray} }
\def\eea{ \end{eqnarray} }

\def\b0{{\bf 0}}

\ifCLASSOPTIONonecolumn
  \renewcommand{\baselinestretch}{1.39}

\else

\fi

\ifCLASSOPTIONonecolumn
\fi

\begin{document}

\title{Analysis, Modeling and Design of Personalized Digital Learning Environment}
\author{ Sanjaya Khanal and Shiva Raj Pokhrel
\thanks{Authors are with
the School of Information Technology,
Deakin University, Geelong, VIC 3220, Australia
(corresponding: shiva.pokhrel@deakin.edu.au)

}
\vspace{-7 mm}}

\maketitle

\begin{abstract}

This research analyzes, models and develops a novel Digital Learning Environment (DLE) fortified by the innovative \textit{Private Learning Intelligence} (PLI) framework. The proposed PLI framework leverages federated machine learning (FL) techniques to \textbf{ }autonomously construct and continuously refine personalized learning models for individual learners, ensuring robust privacy protection. Our approach is pivotal in advancing DLE capabilities, empowering learners to actively participate in personalized real-time learning experiences. The integration of PLI within a DLE also streamlines instructional design and development demands for personalized teaching/learning. 
We seek ways to establish a foundation for the seamless integration of FL into teaching/learning systems, offering a transformative approach to personalized learning in digital environments. Our implementation details and code are made public in \href{https://github.com/pli-research-d/Secure-ML-with-BERT}{GitHub}.
\end{abstract}


\section{Introduction}


The evolving landscape of AI/ML and large language models present a promising avenue for transforming digital learning environments (DLEs)~\cite{brusilovsky2023ai, 10335741} with a strong emphasis on personalized learning for students~\cite{10500382, zhang2020understanding, yannier2024ai}. The integration of personalized learning models, shared (generalized) machine learning models,  interoperable expert knowledge models, recommendation systems, and natural language processing not only tailors educational content to individual student needs but also fosters a more engaging and adaptive learning experience~\cite{aroyo2006interoperability}. Through the application of federated machine learning (FL), privacy concerns in collaborative machine learning are effectively addressed ~\cite{lwakatare2020large}, allowing for the decentralized construction of personalized learning for students. 


Looking forward, the future of DLEs envisions a landscape where AI/ML innovations empower educators to deliver truly personalized and secure learning experiences, ensuring that each student can thrive in their educational pursuits. This paper focuses on personalized learning~\cite{tetzlaff2021developing}, employing a new approach that tailors educational experiences according to individual needs~\cite{zhang2020understanding}, with the potential to positively impact educational outcomes. Personalization is a fundamental attribute of learning experience design, allowing for the adaptation of content and methodologies to meet individual learning styles, preferences, and needs~\cite{ameloot2023supporting}.

\begin{figure}[t]
    \centering
    \includegraphics[scale=0.35]{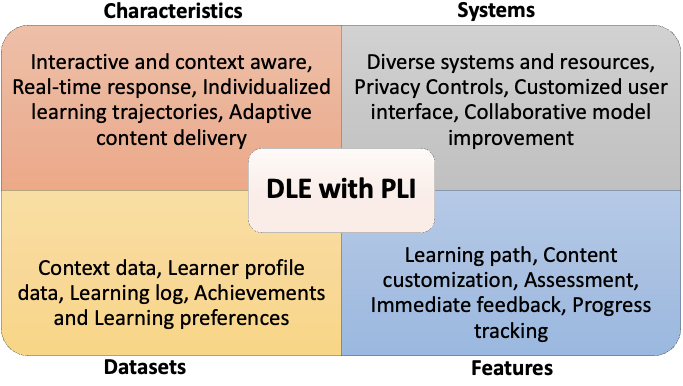}
    \caption{A high-level illustration of the integration and interplay between components of a DLE with PLI, all focused on providing an individualized and interactive learning experience.}\label{fig:fig1}
\end{figure}
This work emphasizes Mayer's Personalization Principle~\cite{mayer2014multimedia}, which asserts that the efficacy of learning is significantly enhanced when it mirrors the dynamics of social discourse, engaging learners in an environment of conversation rather than a passive reception of information. 
We aim to operationalize this personalization approach to promote active cognitive processing, transforming the learning experience into an interactive dialogue.

In contrast to the one-size-fits-all approach in many DLEs~\cite{armatas2003impacts}, in this paper, we aim for modelling private learning intelligence (PLI) to harness the components of a DLE, all focused on providing an individualized
and interactive learning experience, for demonstrating the necessity for real-time personalization in learning. An abstract depiction of how a DLE with PLI works to create a customized and engaging learning journey for each learner is shown in Fig.~\ref{fig:fig1}. While DLEs serve as essential platforms for online learning, they often lack real-time personalization features, relying on the gradual refinement of experiences through historical data analysis ~\cite{maier2022personalized}. In Table~\ref{table-context-of-personalization1}, we explore and compare the contexts \cite{macfadyen2010mining} to illustrate where and how real-time personalization comes into the DLE.

To address the challenges and observations, leveraging big data insights from consumer psychology~\cite{matz2017using} and adopting mass personalization principles~\cite{aheleroff2021mass} has been used to inform scalable, individualized learning paths. Additionally, the integration of Semantic Web technologies in distance learning~\cite{bashir2023systematic} demonstrates the potential for enhancing real-time personalization in DLEs. However, implementing this in practice poses challenges, including time-intensive data analysis, privacy concerns, and consistent interpretation of learner behavior~\cite{duin2020current}, {requiring careful consideration for effective incorporation of collaborative learning into the design and operation of DLE.} Therefore, in this research, we are using DLE data to demonstrate how to develop a \textbf{personalized model}.

\begin{table} [t]
\begin{center}
\begin{tabular}{p{1.9cm}||p{6cm}}
    \hline
        \textbf{DLE components} & \textbf{Context of Personalization}
        \\
        \hline
            Content Search & Searching using natural language and offering
search suggestions\cite{montalvo2018building}
        \\
        \hline
            Content Modules & Suggesting contents with higher/lower difficulty level based on the learner's performance\cite{matsuda2020effect}
        \\
        \hline
            Content Access & Releasing contents based on the learner's current stage in the learning journey\cite{gomede2021deep}
        \\
        \hline
            Content Format & Offering tools for learner to customize contents, and choices in format that suits their preferences\cite{carver1999enhancing}\cite{mayer2014multimedia}
        \\
        \hline
            Discussion \newline Forum & 
            Classifying forum posts based on topics to promote effective discussion forum participation\cite{yang2022untangling}
        \\
        \hline
            Learner support & Chatbot directing learner to help and support resources\cite{fryer2017stimulating}
        \\
        \hline
            Quiz & On-demand knowledge checks quiz from contents and personalized follow-up questions for mastery \cite{minn2022ai}
        \\    
        \hline
            Written \newline Assessments & Automated self-reflection questions to promote self-regulated learning
            \cite{khiat2022self}
        \\
        \hline
            Survey & Prompting learner with automated questionnaire to gauge their current sentiment \cite{prinsloo2019student}\\
        \hline
            Analytical data & Predicting learner performance based on historical data and learning patterns\cite{liao2022deploying}
        \\         
        \hline            
   
\end{tabular}
\end{center}
\caption{Contexts: Where and how real-time Personalization comes into the context of DLE}
\label{table-context-of-personalization1}
\end{table}
\begin{table*}[t]
\caption{Categorization of the State of the Art and Literature on ML-Driven DLE\label{tab:relwork1} }
\centering
\begin{tabular}{p{2.0cm}||p{4cm}p{4.5cm}p{4.5cm}}
\hline
Literature & Learning Management & Personalization & Methodology \\ 
\hline
A. Educational Platform ~\cite{montalvo2018building}~\cite{tapalova2022artificial}~\cite{xu2006intelligent}~\cite{iqbal2023real}
& 
Intelligent Tutoring Systems (ITS); Virtual Learning Environments (VLEs)
&
Adaptive interface for online learners
&
Natural Language Processing (NLP); Clustering Algorithm; Intelligent Agents with Fuzzy Logic
\\
\hline
B. Content ~\cite{diwan2023ai}~\cite{weng2023instructional}~\cite{bezirhan2023automated}
& 
Computer-Based Assessments (CBA)
&
Learning pathway augmentation; Automated Feedback
&
Domain-Specific Automatic Question Generation (AQG) Models; Browser-Based Agents for Collocation Detection; Non-Template-Based Automated Item Generation (AIG)
\\
\hline
C. Teaching ~\cite{dai2022educational}~\cite{dai2023exploring}~\cite{matsuda2020effect}~\cite{mccarthy2018metacognitive}
& 
Digital game-based learning environments (DGBLE); Intelligent Tutoring Systems (ITS)
&
Integrated learning supports; Meta-tutor for personalized guidance; Metacognitive scaffolding
&
Teachable agents; AI-based metacognitive prompts
\\
\hline
D. Learning ~\cite{hwang2020fuzzy}~\cite{tan2022systematic}~\cite{gulz2020preschoolers}
& 
Collaborative Learning Environments; Maths learning game
&
Importance of considering  cognitive and emotional aspects; Differentiated learning materials; Human-computer partnership for learning; Gaze Behaviors and Eye-Tracking for assessing cognitive processes
&
Fuzzy inferencing; Expert systems; Prescriptive AI actions; Teachable agent (TA)
\\
\hline
E. Assessment
~\cite{minn2022ai}~\cite{li2023using}
& 
Intelligent Tutoring Systems (ITS); Cognitive Diagnostic Assessments; Web-based tool for peer review
&
$-$
&
Static and Dynamic Data driven methods in Knowledge Assessment and Peer Review
\\
\hline
F. Feedback
~\cite{huang2023effects}~\cite{maier2022personalized}
& 
Personalized Recommendation Systems; Flipped classrooms and digital learning environment
&
Consideration of students' learning profiles, test results, and video-viewing behavior in providing personalized recommendations; Feedback adaptation for individual student needs
&
SPRT (Sequential Probability Ratio Test) and Bayes' Theorem for assessing students' mastery of concepts
\\
\hline
G. Support
~\cite{fryer2017stimulating}~\cite{gomede2021deep}
& 
Adaptive E-learning Systems (AES); Learning support
&
Features of recommendation systems: relevance, novelty, serendipity, diversity
&
Chatbots; Intelligent tutors and agents; Performance metrics for recommendation systems
\\
\hline
H. Evaluation
~\cite{yang2022untangling}~\cite{liao2022deploying}
& 
$-$
&
Identifying relevant forum posts patterns for individual learner support; Learner autonomy and engagement in self-initiated learning issues
&
Methods for classifying and clustering forum post data; Multimodal data analysis for identifying digital distractions
\\
\hline
\end{tabular}
\end{table*}
AI/ML technologies present a myriad of opportunities~\cite{lwakatare2020large, 9353389}, particularly in the realm of personalizing learners' experiences~\cite{bhutoria2022personalized}. {By leveraging datasets encompassing learner behaviour, including activities like course content visits, time spent, assessment attempts, and scores~\cite{duin2020current}, as well as more intricate behavioural data like sentiment, gaze, physical movements, and environmental factors~\cite{gandhi2023multimodal}, these technologies enable the provision of real-time, personalized recommendations.} Our aim in this work is to seek ways to develop a new framework for integrating large models into the FL for personalization. Such a framework not only caters to learner performance, needs, and preferences but also facilitates continuous monitoring of progress and allows for timely interventions~\cite{chen2020application}. 

{{Table \ref{table-context-of-personalization1} delves into the specifics of where and how real-time personalization intersects with DLEs.}} Despite the recent emergence of large language models such as GPT-4 and Bard, along with public AI tools like ChatGPT, which provide some contextualization in generated content, concerns linger regarding the specificity, validity, and user data privacy associated with their outputs. 
{Instead, a federated network of interoperable, specialized models through standardized protocols and integration can leverage the strengths of domain-specific knowledge for enhanced specialization. These models can incorporate human expert oversight for accuracy and address sustainability by potentially reducing the energy footprint through optimized, smaller models, thereby fostering innovation, efficiency, and accessibility ~\cite{lwakatare2020large}. Additionally, for high personalization, we suggest the development of user-specific models trained on each behavior and action, enabling personalized predictions~\cite{mun2003predicting}. Further, aggregating individual models to construct consolidated models, provides opportunities for continuous development of shared models that are both data-driven and generalized global machine learning models~\cite{lim2020federated}. Such models, when integrated with a federated network of expert knowledge and natural language queries, enhance personalized, credible, and engaging interactions.}

Federated learning, marked by decentralized computation of datasets and model development, is emphasized as a pivotal approach in creating internal AI tools that are both efficient and secure~\cite{lwakatare2020large}. Unlike centralized alternatives, FL ensures a distributed and collaborative approach to model training, aligning with the principles of efficiency and security in the development of AI tools for personalized learning.

With FL, user data remains confined within the device or the DLE application server and is processed in the user's private sandbox environment~\cite{marchiori2022android}. This setup allows for the development and real-time implementation of personalized features specifically tailored for the user, without the need for data to exit the private environment for raw data consolidation. {Multiple ML models can be trained on different datasets so they share their knowledge to improve their overall performance. It is done without exchanging the underlying data, which can be sensitive or privacy-protected. 
This is crucial when dealing with sensitive data, such as learner interactions and behaviours, ensuring privacy and compliance with regulations like GDPR or COPPA~\cite{truong2021privacy}. To resolve the challenges discussed earlier, we have extended and employed the ideas developed for tailored learning in \cite{bellarhmouch2023proposed}, analytics in \cite{mutimukwe2022students} and the personalization \cite{xin2021systematic}.

\subsection{Proposed Idea: originality and benefits}

    DLE with PLI integrates advanced data aggregation and analysis capabilities to ensure privacy, enhancing personalized and effective learning experiences.{ Fig. \ref{fig:fig1} provides an overview of DLE with PLI and its four major components: Characteristics, Systems, Datasets, and Features. The Characteristics ensure interactivity, real-time responses, personalized learning paths, and adaptive content. The Systems include diverse resources, privacy protection, customized interfaces, and collaborative improvement. Datasets consist of context and learner profile information, learning logs, and preference data. Features cover the learning path, content customization, assessments, immediate feedback, and progress tracking, all contributing to an effective digital learning experience. }

\noindent 1. \textbf{Anticipated PLI Framework}. Imagine a world where learning systems are not only intelligent but also respect our privacy. That is precisely what the \textit{Private Learning Intelligence} (PLI) framework envisions. By seamlessly integrating FL with cutting-edge large language models, PLI transforms DLEs into powerful tools for personalized education. The goal is clear: \textit{standardize processes to guarantee consistent model performance across decentralized datasets}. This is not just about improving learning outcomes; it is about safeguarding privacy while maximizing educational potential.

\noindent In Fig.~\ref{fig:fig2}, we showcase the modules within the PLI framework and illustrate how they interact to create a dynamic learning environment. With PLI, personalized learning isn't just a dream—it is a reality that respects your privacy and empowers you to reach your full potential.

\noindent 2. \textbf{Comprehensive Exploration}.
    Within this transformative PLI framework, we unveil visionary proof-of-concept design. 
    Each design represents a leap forward in personalized, privacy-conscious education, poised to revolutionize digital learning environments. As we delve deeper into the PLI framework, our mission becomes clear: \textit{to revolutionize education by prioritizing personalized learning experiences without compromising student privacy}. Gone are the days of one-size-fits-all education; with PLI, every student's journey is tailored to their unique needs and preferences.

\section{Literature}

The literature in this area is quite rich as the current research landscape emphasizes the development of systems tailored to individual learner needs. Diverse methodologies converge on the central theme of enhancing learner experience and efficacy through advanced ML techniques~\cite{montalvo2018building, tapalova2022artificial}. We categorize the exiting work in Tab.~\ref{tab:relwork1} and explain comparatively in the following.

In the domain of educational platforms, as shown in Tab.~\ref{tab:relwork1}, the integration of Natural Language Processing (NLP) and Intelligent Tutoring Systems (ITS) ensures content accessibility and adaptability to learner performance fluctuations~\cite{xu2006intelligent, iqbal2023real}. Intelligent agents in virtual environments foster self-directed learning. Private Learning Intelligence (PLI) provides dynamic adaptation to diverse learner datasets while preserving data privacy~\cite{montalvo2018building, tapalova2022artificial}.

The role of AI in content creation and adaptation, tabulated in the second row of Tab.~\ref{tab:relwork1}, is developed by emphasizing personalized learning pathways~\cite{diwan2023ai, weng2023instructional}. {OpenAI's models can be employed to contribute to personalized assessment through reading passages tailored to individual profiles~\cite{bezirhan2023automated}}. PLI leverages data for effective question set generation and broader Computer-Based Assessments (CBAs); see Fig.~\ref{fig:fig2} for details.

Studies utilize ITSs and teachable agents for nuanced, learner-specific cognitive support~\cite{dai2022educational, matsuda2020effect}. This is shown in the third row of Tab.~\ref{tab:relwork1}. We can see how pedagogical implications of AI-driven metacognitive prompts advance ITS efficacy~\cite{mccarthy2018metacognitive}. Such an approach can be considered into PLI to optimize personalized learning supports in Digital Game-Based Learning Environments (DGBLEs)~\cite{dai2023exploring} while respecting privacy.

Collaborative learning shown in the fourth row of Tab.~\ref{tab:relwork1} represents the call of works that considered cognitive and emotional aspects in AI-facilitated adaptive learning systems is crucial~\cite{hwang2020fuzzy, tan2022systematic, 10367874}. With PLI, we seek to facilitate the creation of nuanced, personalized learning materials based on collective insights from diverse environments~\cite{gulz2020preschoolers}.

AI revolutionizing assessment techniques is another category in Tab.~\ref{tab:relwork1}., enhancing knowledge assessments and feedback accuracy~\cite{minn2022ai, li2023using}. We seek ways for PLI to enhance ITS, providing robust, privacy-preserving personalization beyond NLP capabilities.

AI in feedback is another important class in Tab.~\ref{tab:relwork1} which works as a system to enhance personalized feedback in flipped classrooms~\cite{huang2023effects, maier2022personalized}. We aim to account for this in PLI as recommendation systems by aggregating data for comprehensive personalization; see later in Fig.~\ref{fig:fig2} for details.

We consider support systems and valuation of educational interventions as other key factors in Tab.~\ref{tab:relwork1}. AI augments support systems, emphasizing learner engagement and personalized experiences~\cite{fryer2017stimulating}~\cite{gomede2021deep}. PLI could further enhance AI systems with decentralized, collaborative machine learning while respecting user privacy.
Evaluation of educational interventions using learning analytics advocates for a nuanced understanding of learner autonomy within AI-enriched environments~\cite{yang2022untangling}~\cite{liao2022deploying}. PLI could enable more personalized, context-aware learning experiences while maintaining privacy and security.

Collectively, these studies broaden AI and ML applications in education, providing empirical evidence for effectiveness. A personalized learning framework with distributed learning technologies signifies a learner-centered shift. In other contexts, FL has succeeded in various areas but requires evaluations and addressing limitations in education~\cite{marchiori2022android, brisimi2018federated, pokhrel2020federated}.

\section{Proposed PLI Framework}
In the realm of education technology, the evolution of DLEs has been marked by a quest for enhanced personalization \cite{bellarhmouch2023proposed, carver1999enhancing} and privacy \cite{prinsloo2019student, mutimukwe2022students}. In this work, we present a significant advancement—a model-driven personalized DLE, which serves as the cornerstone for the introduction of a groundbreaking framework we term ``Private Learning Intelligence" (PLI). This innovative framework takes inspiration from the Android Private Core Computing (PCC) architecture~\cite{marchiori2022android}, and its core principle revolves around the utilization of federated learning techniques to construct machine learning models while upholding stringent privacy safeguards \cite{lim2020federated}.


 
\begin{figure}[t]
    \centering
    \includegraphics[scale=0.345]{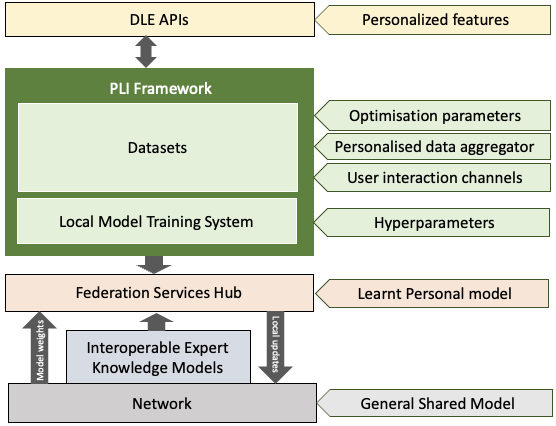}
    \caption{Architecture of the Private Learning Intelligence (PLI) for personalized learning in DLE}\label{fig:fig2}
\end{figure}

We enter the realm of PLI, where innovation meets the necessity to address two critical challenges head-on. Firstly, we tackle the imperative for dynamic, locally-trained personal ML model training, safeguarding sensitive data within the user's environment. No longer will privacy concerns hinder progress; PLI ensures our data stays where it belongs—under our control.

PLI goes beyond mere privacy protection; it pioneers the seamless integration of {local personal models and global knowledge models}. Imagine unlocking a world of personalized, domain-specific insights directly within our learning experience. With PLI, education becomes more than just acquiring knowledge—it's about harnessing insights tailored to our unique journey and aspirations. We aim to revolutionize learning through the power of PLI, where privacy and personalization converge to shape a brighter future.


\begin{figure}
    \centering \includegraphics[scale=0.159]{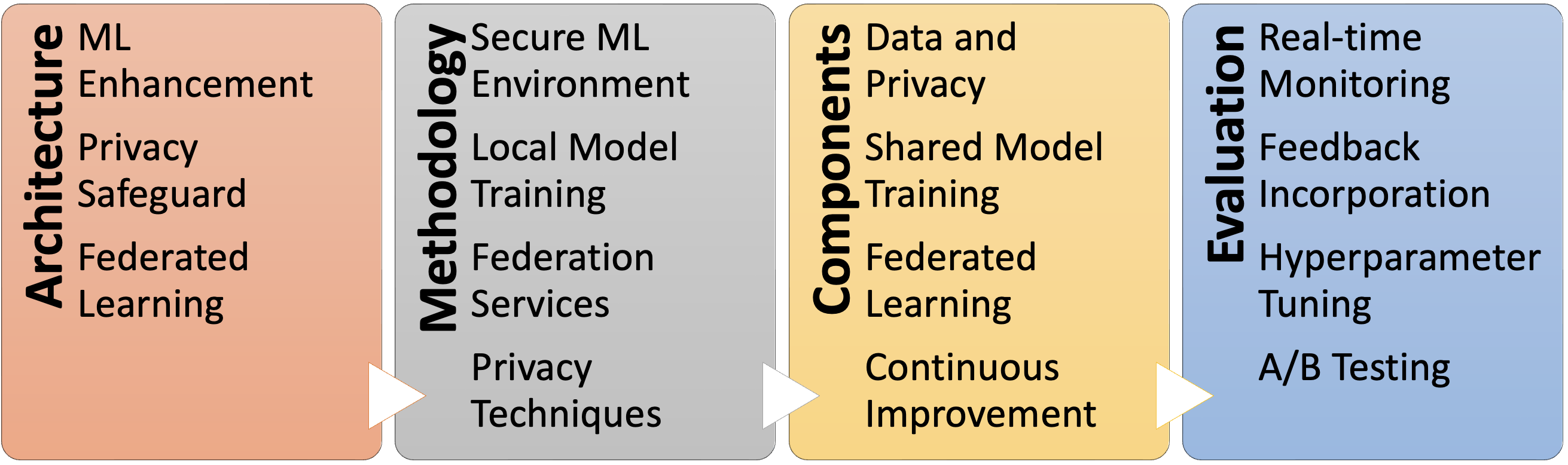}
    \caption{Four-stage PLI implementation process, focusing on federated learning enhancement, privacy assurance, secure local model training, collaborative improvement, and performance optimization through real-time monitoring and testing}\label{fig:fig3}
\end{figure}

\subsection{Methodology}
PLI incorporates sandboxing techniques to establish a secure, isolated ML environment within the user device or DLE server by design. Utilizing federated learning, it allows participants to train local models using their data, while only sharing non-sensitive local updates with the global model. This approach minimizes data privacy concerns as highlighted by Duy et al. (2021)\cite{duy2021confidential}. The model is designed to learn and improve over time, leveraging local user data without violating privacy norms.

Fig. \ref{fig:fig3} illustrates the process involved in implementation PLI, segmented into four main stages. The architecture includes enhancing machine learning for federated contexts, ensuring privacy, and the foundational concept of federated learning. The methodology involves creating a secure environment, training models locally, coordinating nodes through federation services, and using privacy techniques like differential privacy. Components comprise data and privacy considerations, collaborative model training, continuous system improvement, and the central role of federated learning. Evaluation entails real-time monitoring, feedback incorporation, hyperparameter tuning, and A/B testing to optimize model performance.


\subsection{Architecture and Features}
 
{As mentioned, Fig. \ref{fig:fig2} outlines the architecture of a personalized learning infrastructure (PLI) for machine learning, featuring a PLI Framework as the core component that governs the system, including algorithms and protocols for personalization. It incorporates a Local Model Training System for training machine learning models on a local scale, using datasets provided or accessed via the PLI Framework. Datasets are tailored to individual users, and DLE APIs allow external systems to interact with the PLI. Personalized Features, Optimization Parameters, and Hyperparameters are used to tailor the system and optimize model performance. A Personalized Data Aggregator collects data from user interactions, while User Interaction Channels provide avenues for interaction. This architecture delegates the local model for personalizing interactions with individual users or their tasks, the General Shared Model for understanding the broader applicability of services, and offers integration capabilities with third-party interoperable expert knowledge models for enhancing personalization with expert knowledge.} 

\subsection{Components}
PLI can be viewed as a secure and isolated (sandboxed) ML environment in the user device/edge server by design with features running computations on user data. Results are surfaced through the UI or open-source APIs.
It enables ML locally using the user data and processes from the app while maintaining separation from the rest of the app. App permissions are used to restrict access to its datasets and the Local Model Training System (LMTS). Only local updates that exclude users’ private information are sent externally through the Federation Services Hub to the Network for global model aggregation, and then new model weights are received as updates for the LTMS. The aim is to enhance DLE personalization features over time by learning from the data.
   
\begin{figure}
    \centering
    \includegraphics[scale=0.637]{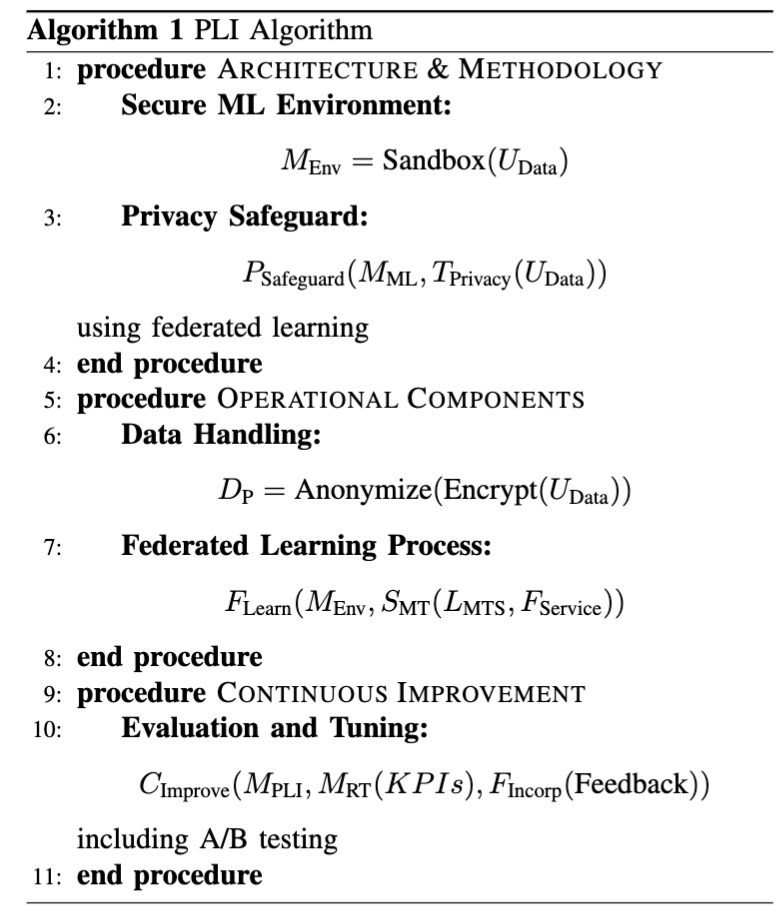}
    \label{algorithm1}
\end{figure}
\begin{table*}[t]
\begin{center}
\begin{tabular}{p{2.5cm}||p{3.5cm}||p{2.15cm}||p{3cm}||p{4cm}}
    \hline
        \textbf{DLE tracking data variables} & \textbf{Learner behavior indicators} & \textbf{Learning measures and scores} & \textbf{Time series analysis} & \textbf{Example of personalization strategies }
        \\
        \hline        
            Logins 
            & $D$ Streaks days: [1]=Low(2); [2]=Moderate(4); [3]=High(6); [4,7]=Exceptional(10)
            & Conscientiousness [2, 4, 6, 10]
            & Analyze login streaks to gauge commitment
            & Encourage learners to maintain streaks; Provide support resources if progress is lagging
        \\
        \hline
            Time Spent 
            & $T$ Time spent (hours): [1,3]=Low(2); [4,7]=Moderate(4); [8,12]=High(6); [13,20]=Exceptional(10)
            & Conscientiousness [2, 4, 6, 10]
            & Average time between logins to gauge regularity
            &  Offer support based on engagement patterns; Tailor notifications to encourage consistent learning
        \\
        \hline
            Page Visits 
            & $P$ Page visit count: [1]=Low(2); [2]=Moderate(4); [3]=High(6); [4+]=Exceptional(10)
            & Motivation [2, 4, 6, 10]
            & Track content exploration to understand preferences
            & Customize content and structure to match learning styles; Recommend content based on real-time data.
        \\
        \hline
            Search Queries 
            & $S$ Search queries count: [1]=Low(2); [2,3]=Moderate(4); [4,5]=High(6); [6+]=Exceptional(10)
            & Motivation [2, 4, 6, 10]
            & Analyze searches to infer needs and context  
            & Improve search relevance by adjusting results to align with learner expectations
        \\
        \hline
            Activity Completion
            & $C$ Activity completion [1\%,20\%]=Low(2); [21\%,60\%]=Moderate(4); [61\%,90\%]=High(6); [91\%,99\%]=Very High(8); [100\%]=Exceptional(10)
            & Understanding [2, 4, 6, 8, 10]
            & Monitor progress in activities  
            & Adjust interactive content difficulty and type based on responses to match ability and style
        \\
        \hline
            Quiz Performance
            & $Q$ Average Quiz score: [1\%,50\%]=Low(2); [51\%,70\%]=Moderate(4); [71\%,90\%]=High(6); [91\%,99\%]=Very High(8); [100\%]=Exceptional(10)
            & Understanding [2, 4, 6, 8, 10]
            & Average score improvement over time assess learning progression
            & Provide customized feedback and targeted questions to reinforce concepts
        \\    
        \hline
            Content Reactions 
            & $R$ Content reactions ratio (Positive:Negative): [1:1]=Low(2); [2:1]=Moderate(4); [4:1]=High(6); [7:1]=Exceptional(10)
            & Engagement [2, 4, 6, 10]
            & Determine the learner's emotional response to content
            & Direct learners to content format that  have elicited positive engagement or their preferred learning style
        \\         
        \hline
            Feedback 
            & $F$ Satisfaction score average: [1,3]= Low(2); [4,5]=Moderate(4); [6,7]=High(6); [8,10]=Exceptional(10)
            & Engagement [2, 4, 6, 10]
            & Analyze direct feedback to gauge satisfaction
            & Correlate satisfaction with engagement metrics to offer adaptive communication and further exploration of support options.
        \\ 
\end{tabular}
\end{center}
\caption{Proposed classification of real-time learning measures for the anticipated personalization in the developed PLI framework}
\label{table-participation data and PLI-2}
\end{table*}
{
\subsubsection{Interfaces Design}
The Application Programming Interface (API) serves as a vital link facilitating communication between PLI and diverse components of DLEs, allowing for the exchange of context and content data. Datasets from DLE components train the \textbf{local personalized model}, ensuring context-aware solutions, while user inputs gathered by the PLI User Interface (UI) undergo content analysis, enhancing content-awareness through queries to expert knowledge models. Both the API and UI are essential for personalization.

The UI collects user inputs for content analysis, utilizing topic modeling algorithms and semantic extraction to identify and categorize themes and extract significant terms. \textbf{Personalized model} customizes this process, adjusting the importance of topics and keywords based on user interactions. These insights enable targeted searches through expert knowledge models, directly addressing content. The UI also facilitates user feedback, refining algorithms and enhancing precision over time.

\subsubsection{User Data and Privacy}
Personalization using PLI can be achieved by tracking data variables for each learner {which} are utilized for the construction of a model. These aspects are illustrated with discussions in Table \ref{table-participation data and PLI-2}. The {local personalized model} begins as a general global model and is subsequently retrained. To optimize model parameters and prevent overfitting, cross-validation techniques are employed, ensuring the model's applicability.

Standardized values represent the average interactions of a learner with the DLE to analyze their engagement and performance over a period. For each {variable i} the first column of Table \ref{table-participation data and PLI-2}, learner behavior indicators (second column of Table \ref{table-participation data and PLI-2}) are utilized to ascertain corresponding learning measure scores. The computed measures and scores (third column of Table \ref{table-participation data and PLI-2} indicate how each feature influences the prediction of whether a learner is a high or a low performer.

Learning measures are ascertained through statistical analysis and clustering techniques such as K-means or hierarchical clustering. Considering this context in vectored form, for the anticipated multilevel classification, our objective is to perform a time series analysis referencing the learning measure score to create a comprehensive set of personalized insights (corresponding to the learning measures vector) that reflect various dimensions of the learner's interactions and outcomes.\footnote{Personalization strategies that can be developed from the insights are listed as examples only and need further investigation in future.}

Next, we consider an example where learner login data is analyzed to calculate the average time between logins, and quiz results are evaluated to determine the average score improvement over time. These average metrics provide a foundation for understanding learner engagement and learning progress, respectively. Such a simple analysis can reveal the existence of different learner clusters (as mentioned earlier), such as highly engaged learners and those facing challenges in their learning process. To this end, a highly engaged learner characterized by consistent improvement, could be offered more complex content and advanced subjects to maintain their interest and deepen their learning. Conversely, a learner with irregular login patterns and no significant improvement in quiz scores may require access to basic resources and frequent, targeted interventions to address their learning gaps.}

PLI is designed and developed taking into account user privacy. We employ the FL framework to improve underlying ML models, so no sensitive user data leaves the local device. Note that the access permissions are strictly regulated to protect datasets. Furthermore, user data can be anonymized and encrypted before parsing for the training. Strict role-based access controls can also be implemented to prevent unauthorized access to the data. Data will then be primarily stored locally and only essential, non-private information will be sent to the cloud for model updates. Before training, all sensitive data shall be anonymized or encrypted. The model can then help by generating summaries that never include sensitive information. Moving beyond, techniques like differential privacy could be applied for data sanitization.

\subsubsection {Learner Personalized Model}
Our implementation involved fine-tuning the base {llama2-7b-chat model}\footnote{https://ollama.com/library/llama2:7b-chat} on a specifically curated dataset as learner behaviour indicators for identifying learning
measures (Table \ref{table-participation data and PLI-2}), thereby learning correlations between input text and student's learning characteristics.
\begin{figure}
    \centering
    \includegraphics[scale=0.67]{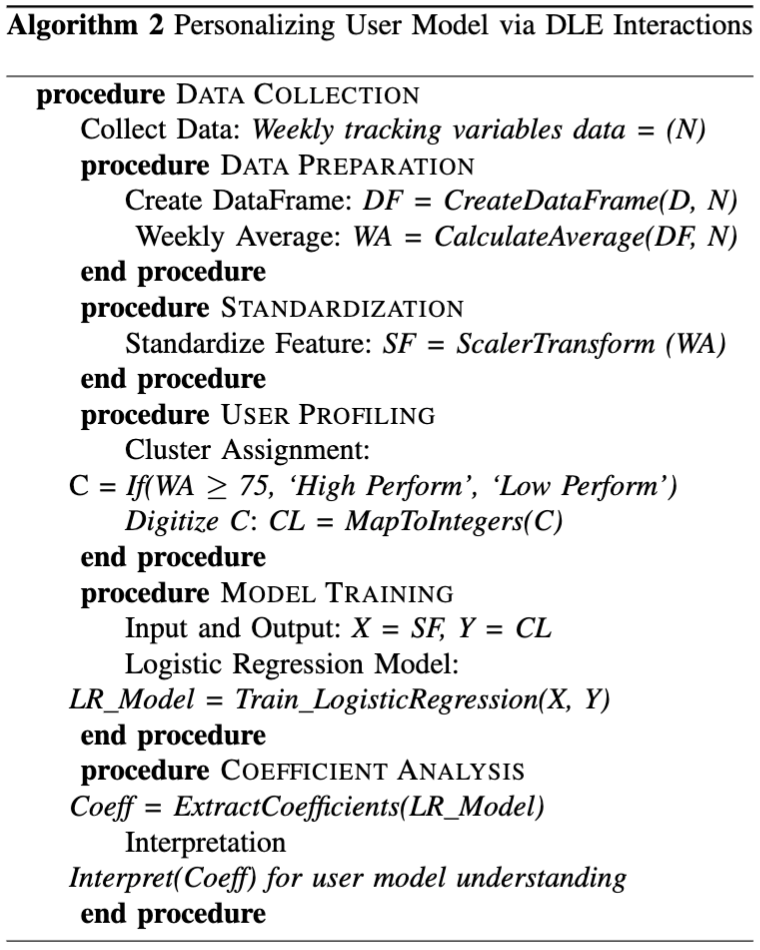}
    \label{algo:perDLE}
\end{figure}
The model training shown in Algorithm~\ref{algo:perDLE} involves standardizing weekly data and training a predictive model to classify users as ``High Performer" or ``Low Performer" based on average scores. This classification is quantified by mapping cluster labels to integers for model training. A logistic regression model is used for testing, which highlights the significance of various features in predicting user performance. The process also includes standardizing daily data to train another logistic regression model, which provides an accuracy metric and insights into the significance of daily activity features in the {local personalized model}.

The process of periodic retraining shown in Fig. \ref{fig:fig4} involves the following steps: initial training of a global model on local devices, distribution of this model by a central server to all clients, local updates of the model by clients using their own data, aggregation of these updates by the central server to improve the global model, and redistribution of the updated model to clients for further training. This cycle repeats periodically, ensuring the model stays relevant and up-to-date. Throughout this process, measures are taken to ensure efficiency, privacy, and security, such as model compression, secure transmission protocols, differential privacy, and continuous monitoring of model performance.

\subsubsection {Federation Services Hub}
The Federation Services Hub runs distributed services and models within the PLI, coordinating interactions between local model training systems and a network of interoperable models. It ensures interoperability, allowing seamless communication and data exchange among different learning models and services. On-device learning emphasizes privacy and efficiency, as raw data does not need to be shared with a central server or other devices, and learning is tailored to the specific context of the device. Local models interact through Federation Services to contribute to the global model. Only model parameters (not the raw data) are sent to the server, ensuring data never leaves the device. The hub also coordinates the interaction between local model training systems and a network of interoperable models. This enables the system to leverage specialized knowledge from various sources and provide content recommendation and adaptive learning pathways to users.
\newline

\subsubsection {Network ML Servers}

The network in the provided architecture plays a crucial role in the PLI by connecting various expert knowledge models for expertise sharing, enhancing personalization, and facilitating scalable learning solutions. The network enables the sharing of specialized expertise and allows the system to cover a broad range of subjects and learning styles. It provides highly personalized learning experiences by tapping into diverse expert models, each contributing its own insights to form a tailored educational path that adapts to the unique needs and preferences of the learner. The network's inherently scalable design ensures that the PLI remains current and can continuously evolve with emerging educational trends and technologies. Interoperable expert knowledge models can be updated independently, allowing for continuous improvement without overhauling the entire system. The process involves a feedback loop where a central server enhances a global model with insights from various LMTSs and distributes to new users as a standard general model. A central ML server aggregates updates from different PLI instances which uses consensus mechanism to ensure that the aggregated model is improved by considering updates from multiple sources. All communication between PLI-integrated DLEs and ML server during the aggregation process is encrypted. The ongoing process of updating machine learning models with new data over time is crucial for maintaining the relevance and accuracy of both local and global models within a DLE.  

\begin{figure}
    \centering \includegraphics[scale=0.207]{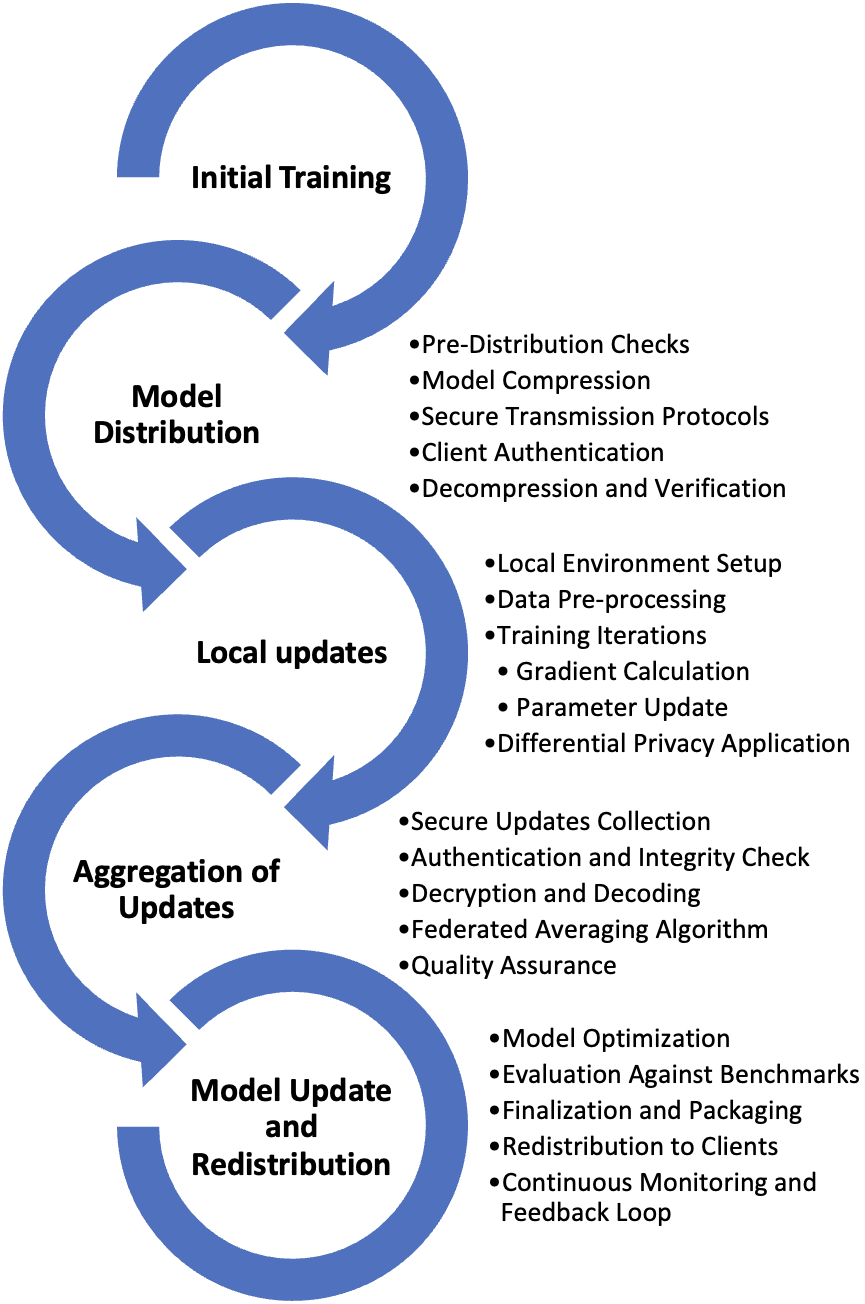}
    \caption{Process of periodic retraining of the models with the latest data to keep them up-to-date}\label{fig:fig4}
\end{figure}  

The architecture is designed to evolve and the implementation details are made public~\href{https://github.com/pli-research-d/Secure-ML-with-BERT}{GitHub}. Metrics are continuously assessed, and we observed that the model adapts to offer increasingly personalized and effective educational experiences. Effective management of machine learning models includes real-time monitoring of key performance indicators (KPIs), incorporating user feedback for model fine-tuning, and regular hyperparameter tuning. A/B testing is used to compare different model versions, while resilience tests assess the model's adaptability to changing data patterns and user behaviours. Periodic assessments evaluate accuracy, bias, and relevance to user needs, and dynamic updating mechanisms facilitate frequent retraining and data integration. Further, addressing the need for continuous training and long-term evaluation can include evaluating the improvements in model performance after each update cycle, conducting longitudinal studies to assess the impact of the feedback loop and continuous training on user engagement, satisfaction, and the achievement of desired outcomes, and continually assessing the privacy implications of the feedback loop. As models are updated with user data, it's crucial to ensure that privacy is not compromised.

\section{Conclusion}
The proposed PLI framework represents a groundbreaking approach in the realm of educational technology, addressing the dual imperatives of enhanced personalization and robust privacy safeguards. It works to augment the educational process by alleviating the need for routine instructional design tasks, thereby enabling teachers to focus on more nuanced aspects of teaching. Further, PLI-integrated DLE enhances student learning by creating engaging and interactive experiences. 
A flipped or blended learning model, which focuses equally on individual progress and communal learning, is recommended in educational settings to benefit from its full potential. There are also implementation challenges, user training, and support structures necessary for the successful deployment of such a system. However, as technology matures, solutions to these issues are expected to emerge, making PLI a cornerstone in the development of next-generation digital learning environment and learning management system. Future research could improve capabilities by analyzing sensor data in smartphones, tablets, and wearable devices, and incorporating the analysis of learners' emotional and cognitive states, to provide a more comprehensive personalized learning experience.

\ifCLASSOPTIONcaptionsoff
  \newpage
\fi

\bibliographystyle{ieeetr}
\bibliography{references}

\begin{thebibliography}{10}

\bibitem{brusilovsky2023ai}
P.~Brusilovsky, ``{AI} in education, learner control, and human-ai
  collaboration,'' {\em International Journal of Artificial Intelligence in
  Education}, pp.~1--14, 2023.

\bibitem{10335741}
D.~Roldán-Álvarez and F.~J. Mesa, ``Intelligent deep-learning tutoring system
  to assist instructors in programming courses,'' {\em IEEE Transactions on
  Education}, vol.~67, no.~1, pp.~153--161, 2024.

\bibitem{10500382}
F.~Fruett, F.~P. Barbosa, S.~C.~Z. Fraga, and P.~I.~A. Guimarães, ``Empowering
  steam activities with artificial intelligence and open hardware: The
  bitdoglab,'' {\em IEEE Transactions on Education}, pp.~1--10, 2024.

\bibitem{zhang2020understanding}
L.~Zhang, J.~D. Basham, and S.~Yang, ``Understanding the implementation of
  personalized learning: A research synthesis,'' {\em Educational Research
  Review}, vol.~31, p.~100339, 2020.

\bibitem{yannier2024ai}
N.~Yannier, S.~E. Hudson, H.~Chang, and K.~R. Koedinger, ``{AI} adaptivity in a
  mixed-reality system improves learning,'' {\em International Journal of
  Artificial Intelligence in Education}, pp.~1--18, 2024.

\bibitem{aroyo2006interoperability}
L.~Aroyo, P.~Dolog, G.-J. Houben, M.~Kravcik, A.~Naeve, M.~Nilsson, and
  F.~Wild, ``Interoperability in personalized adaptive learning,'' {\em Journal
  of Educational Technology \& Society}, vol.~9, no.~2, pp.~4--18, 2006.

\bibitem{lwakatare2020large}
L.~E. Lwakatare, A.~Raj, I.~Crnkovic, J.~Bosch, and H.~H. Olsson, ``Large-scale
  machine learning systems in real-world industrial settings: A review of
  challenges and solutions,'' {\em Information and software technology},
  vol.~127, p.~106368, 2020.

\bibitem{tetzlaff2021developing}
L.~Tetzlaff, F.~Schmiedek, and G.~Brod, ``Developing personalized education: A
  dynamic framework,'' {\em Educational Psychology Review}, vol.~33,
  pp.~863--882, 2021.

\bibitem{ameloot2023supporting}
E.~Ameloot, R.~Tijs, A.~Thomas, B.~Rienties, and T.~Schellens, ``Supporting
  students’ basic psychological needs and satisfaction in a blended learning
  environment through learning analytics,'' {\em Computers \& Education},
  p.~104949, 2023.

\bibitem{mayer2014multimedia}
R.~E. Mayer, ``Multimedia instruction,'' {\em Handbook of research on
  educational communications and technology}, pp.~385--399, 2014.

\bibitem{armatas2003impacts}
C.~Armatas, D.~Holt, and M.~Rice, ``Impacts of an online-supported,
  resource-based learning environment: Does one size fit all?,'' {\em Distance
  education}, vol.~24, no.~2, pp.~141--158, 2003.

\bibitem{maier2022personalized}
U.~Maier and C.~Klotz, ``Personalized feedback in digital learning
  environments: Classification framework and literature review,'' {\em
  Computers and Education: Artificial Intelligence}, vol.~3, p.~100080, 2022.

\bibitem{macfadyen2010mining}
L.~P. Macfadyen and S.~Dawson, ``Mining lms data to develop an “early warning
  system” for educators: A proof of concept,'' {\em Computers \& education},
  vol.~54, no.~2, pp.~588--599, 2010.

\bibitem{matz2017using}
S.~C. Matz and O.~Netzer, ``Using big data as a window into consumers’
  psychology,'' {\em Current opinion in behavioral sciences}, vol.~18,
  pp.~7--12, 2017.

\bibitem{aheleroff2021mass}
S.~Aheleroff, N.~Mostashiri, X.~Xu, and R.~Y. Zhong, ``Mass personalisation as
  a service in industry 4.0: A resilient response case study,'' {\em Advanced
  Engineering Informatics}, vol.~50, p.~101438, 2021.

\bibitem{bashir2023systematic}
F.~Bashir and N.~F. Warraich, ``Systematic literature review of semantic web
  for distance learning,'' {\em Interactive Learning Environments}, vol.~31,
  no.~1, pp.~527--543, 2023.

\bibitem{duin2020current}
A.~H. Duin and J.~Tham, ``The current state of analytics: Implications for
  learning management system (lms) use in writing pedagogy,'' {\em Computers
  and Composition}, vol.~55, p.~102544, 2020.

\bibitem{montalvo2018building}
S.~Montalvo, J.~Palomo, and C.~de~la Orden, ``Building an educational platform
  using nlp: A case study in teaching finance.,'' {\em J. Univers. Comput.
  Sci.}, vol.~24, no.~10, pp.~1403--1423, 2018.

\bibitem{matsuda2020effect}
N.~Matsuda, W.~Weng, and N.~Wall, ``The effect of metacognitive scaffolding for
  learning by teaching a teachable agent,'' {\em International Journal of
  Artificial Intelligence in Education}, vol.~30, pp.~1--37, 2020.

\bibitem{gomede2021deep}
E.~Gomede, R.~M. de~Barros, and L.~de~Souza~Mendes, ``Deep auto encoders to
  adaptive e-learning recommender system,'' {\em Computers and education:
  Artificial intelligence}, vol.~2, p.~100009, 2021.

\bibitem{carver1999enhancing}
C.~A. Carver, R.~A. Howard, and W.~D. Lane, ``Enhancing student learning
  through hypermedia courseware and incorporation of student learning styles,''
  {\em IEEE transactions on Education}, vol.~42, no.~1, pp.~33--38, 1999.

\bibitem{yang2022untangling}
B.~Yang, H.~Tang, L.~Hao, and J.~R. Rose, ``Untangling chaos in discussion
  forums: A temporal analysis of topic-relevant forum posts in moocs,'' {\em
  Computers \& Education}, vol.~178, p.~104402, 2022.

\bibitem{fryer2017stimulating}
L.~K. Fryer, M.~Ainley, A.~Thompson, A.~Gibson, and Z.~Sherlock, ``Stimulating
  and sustaining interest in a language course: An experimental comparison of
  chatbot and human task partners,'' {\em Computers in Human Behavior},
  vol.~75, pp.~461--468, 2017.

\bibitem{minn2022ai}
S.~Minn, ``Ai-assisted knowledge assessment techniques for adaptive learning
  environments,'' {\em Computers and Education: Artificial Intelligence},
  vol.~3, p.~100050, 2022.

\bibitem{khiat2022self}
H.~Khiat and S.~Vogel, ``A self-regulated learning management system: Enhancing
  performance, motivation and reflection in learning,'' {\em Journal of
  University Teaching \& Learning Practice}, vol.~19, no.~2, pp.~43--59, 2022.

\bibitem{prinsloo2019student}
P.~Prinsloo, S.~Slade, and M.~Khalil, ``Student data privacy in moocs: A
  sentiment analysis,'' {\em Distance Education}, vol.~40, no.~3, pp.~395--413,
  2019.

\bibitem{liao2022deploying}
C.-H. Liao and J.-Y. Wu, ``Deploying multimodal learning analytics models to
  explore the impact of digital distraction and peer learning on student
  performance,'' {\em Computers \& Education}, vol.~190, p.~104599, 2022.

\bibitem{tapalova2022artificial}
O.~Tapalova and N.~Zhiyenbayeva, ``Artificial intelligence in education: Aied
  for personalised learning pathways.,'' {\em Electronic Journal of
  e-Learning}, vol.~20, no.~5, pp.~639--653, 2022.

\bibitem{xu2006intelligent}
D.~Xu and H.~Wang, ``Intelligent agent supported personalization for virtual
  learning environments,'' {\em Decision Support Systems}, vol.~42, no.~2,
  pp.~825--843, 2006.

\bibitem{iqbal2023real}
M.~Z. Iqbal and A.~G. Campbell, ``Real-time hand interaction and self-directed
  machine learning agents in immersive learning environments,'' {\em Computers
  \& Education: X Reality}, vol.~3, p.~100038, 2023.

\bibitem{diwan2023ai}
C.~Diwan, S.~Srinivasa, G.~Suri, S.~Agarwal, and P.~Ram, ``Ai-based learning
  content generation and learning pathway augmentation to increase learner
  engagement,'' {\em Computers and Education: Artificial Intelligence}, vol.~4,
  p.~100110, 2023.

\bibitem{weng2023instructional}
X.~Weng and T.~K. Chiu, ``Instructional design and learning outcomes of
  intelligent computer assisted language learning: Systematic review in the
  field,'' {\em Computers and Education: Artificial Intelligence}, p.~100117,
  2023.

\bibitem{bezirhan2023automated}
U.~Bezirhan and M.~von Davier, ``Automated reading passage generation with
  openai's large language model,'' {\em arXiv preprint arXiv:2304.04616}, 2023.

\bibitem{dai2022educational}
C.-P. Dai and F.~Ke, ``Educational applications of artificial intelligence in
  simulation-based learning: A systematic mapping review,'' {\em Computers and
  Education: Artificial Intelligence}, p.~100087, 2022.

\bibitem{dai2023exploring}
C.-P. Dai, F.~Ke, Y.~Pan, and Y.~Liu, ``Exploring students’ learning support
  use in digital game-based math learning: A mixed-methods approach using
  machine learning and multi-cases study,'' {\em Computers \& Education},
  vol.~194, p.~104698, 2023.

\bibitem{mccarthy2018metacognitive}
K.~S. McCarthy, A.~D. Likens, A.~M. Johnson, T.~A. Guerrero, and D.~S.
  McNamara, ``Metacognitive overload!: Positive and negative effects of
  metacognitive prompts in an intelligent tutoring system,'' {\em International
  Journal of Artificial Intelligence in Education}, vol.~28, pp.~420--438,
  2018.

\bibitem{hwang2020fuzzy}
G.-J. Hwang, H.-Y. Sung, S.-C. Chang, and X.-C. Huang, ``A fuzzy expert
  system-based adaptive learning approach to improving students’ learning
  performances by considering affective and cognitive factors,'' {\em Computers
  and Education: Artificial Intelligence}, vol.~1, p.~100003, 2020.

\bibitem{tan2022systematic}
S.~C. Tan, A.~V.~Y. Lee, and M.~Lee, ``A systematic review of artificial
  intelligence techniques for collaborative learning over the past two
  decades,'' {\em Computers and Education: Artificial Intelligence}, p.~100097,
  2022.

\bibitem{gulz2020preschoolers}
A.~Gulz, L.~Londos, and M.~Haake, ``Preschoolers’ understanding of a
  teachable agent-based game in early mathematics as reflected in their gaze
  behaviors--an experimental study,'' {\em International Journal of Artificial
  Intelligence in Education}, vol.~30, pp.~38--73, 2020.

\bibitem{li2023using}
A.~W. Li, ``Using peerceptiv to support ai-based online writing assessment
  across the disciplines,'' {\em Assessing Writing}, vol.~57, p.~100746, 2023.

\bibitem{huang2023effects}
A.~Y. Huang, O.~H. Lu, and S.~J. Yang, ``Effects of artificial
  intelligence--enabled personalized recommendations on learners’ learning
  engagement, motivation, and outcomes in a flipped classroom,'' {\em Computers
  \& Education}, vol.~194, p.~104684, 2023.

\bibitem{9353389}
L.~Meng, W.~Zhang, Y.~Chu, and M.~Zhang, ``Ld–lp generation of personalized
  learning path based on learning diagnosis,'' {\em IEEE Transactions on
  Learning Technologies}, vol.~14, no.~1, pp.~122--128, 2021.

\bibitem{bhutoria2022personalized}
A.~Bhutoria, ``Personalized education and artificial intelligence in the united
  states, china, and india: A systematic review using a human-in-the-loop
  model,'' {\em Computers and Education: Artificial Intelligence}, vol.~3,
  p.~100068, 2022.

\bibitem{gandhi2023multimodal}
A.~Gandhi, K.~Adhvaryu, S.~Poria, E.~Cambria, and A.~Hussain, ``Multimodal
  sentiment analysis: A systematic review of history, datasets, multimodal
  fusion methods, applications, challenges and future directions,'' {\em
  Information Fusion}, vol.~91, pp.~424--444, 2023.

\bibitem{chen2020application}
X.~Chen, H.~Xie, D.~Zou, and G.-J. Hwang, ``Application and theory gaps during
  the rise of artificial intelligence in education,'' {\em Computers and
  Education: Artificial Intelligence}, vol.~1, p.~100002, 2020.

\bibitem{mun2003predicting}
Y.~Y. Mun and Y.~Hwang, ``Predicting the use of web-based information systems:
  self-efficacy, enjoyment, learning goal orientation, and the technology
  acceptance model,'' {\em International journal of human-computer studies},
  vol.~59, no.~4, pp.~431--449, 2003.

\bibitem{lim2020federated}
W.~Y.~B. Lim, N.~C. Luong, D.~T. Hoang, Y.~Jiao, Y.-C. Liang, Q.~Yang,
  D.~Niyato, and C.~Miao, ``Federated learning in mobile edge networks: A
  comprehensive survey,'' {\em IEEE Communications Surveys \& Tutorials},
  vol.~22, no.~3, pp.~2031--2063, 2020.

\bibitem{marchiori2022android}
E.~Marchiori, S.~de~Haas, S.~Volnov, R.~Falcon, R.~Pinto, and M.~Zamarato,
  ``Android private compute core architecture,'' {\em arXiv preprint
  arXiv:2209.10317}, 2022.

\bibitem{truong2021privacy}
N.~Truong, K.~Sun, S.~Wang, F.~Guitton, and Y.~Guo, ``Privacy preservation in
  federated learning: An insightful survey from the gdpr perspective,'' {\em
  Computers \& Security}, vol.~110, p.~102402, 2021.

\bibitem{bellarhmouch2023proposed}
Y.~Bellarhmouch, A.~Jeghal, H.~Tairi, and N.~Benjelloun, ``A proposed
  architectural learner model for a personalized learning environment,'' {\em
  Education and Information Technologies}, vol.~28, no.~4, pp.~4243--4263,
  2023.

\bibitem{mutimukwe2022students}
C.~Mutimukwe, O.~Viberg, L.-M. Oberg, and T.~Cerratto-Pargman, ``Students'
  privacy concerns in learning analytics: Model development,'' {\em British
  Journal of Educational Technology}, vol.~53, no.~4, pp.~932--951, 2022.

\bibitem{xin2021systematic}
N.~S. Xin, A.~S. Shibghatullah, M.~H. Abd~Wahab, {\em et~al.}, ``A systematic
  review for online learning management system,'' in {\em Journal of Physics:
  Conference Series}, vol.~1874, p.~012030, IOP Publishing, 2021.

\bibitem{10367874}
N.~Dehbozorgi and M.~T. Kunuku, ``Exploring the influence of emotional states
  in peer interactions on students’ academic performance,'' {\em IEEE
  Transactions on Education}, pp.~1--8, 2023.

\bibitem{brisimi2018federated}
T.~S. Brisimi, R.~Chen, T.~Mela, A.~Olshevsky, I.~C. Paschalidis, and W.~Shi,
  ``Federated learning of predictive models from federated electronic health
  records,'' {\em International journal of medical informatics}, vol.~112,
  pp.~59--67, 2018.

\bibitem{pokhrel2020federated}
S.~R. Pokhrel and J.~Choi, ``Federated learning with blockchain for autonomous
  vehicles: Analysis and design challenges,'' {\em IEEE Transactions on
  Communications}, vol.~68, no.~8, pp.~4734--4746, 2020.

\bibitem{duy2021confidential}
K.~D. Duy, T.~Noh, S.~Huh, and H.~Lee, ``Confidential machine learning
  computation in untrusted environments: A systems security perspective,'' {\em
  IEEE Access}, vol.~9, pp.~168656--168677, 2021.

\end{thebibliography}

\end{document}